# Mechanical properties of ultra-hard nanocrystalline cubic boron nitride

Vladimir L. Solozhenko,[a,*]  Volodymyr Bushlya [b]  and  Jinming Zhou [b]

[a] LSPM–CNRS, Université Paris Nord, 93430 Villetaneuse, France

[b] Division of Production and Materials Engineering, Lund University, 22100 Lund, Sweden

Nanostructure and mechanical properties of bulk nanocristalline cubic boron nitride have been studied by transmission electron microscopy, and micro- and nanoindentation. The obtained data on hardness, elastic properties and fracture toughness clearly indicate that nano-cBN belongs to a family of advanced ultra-hard materials.

*Keywords*:  boron nitride, nanocrystalline, hardness, elastic moduli, fracture toughness.

**Introduction**

Cubic (F-43m) boron nitride (cBN) is a superhard phase that is about half as hard as diamond [1] but with much higher thermal and chemical stability [2]. The latter makes cBN a material of choice for wide range of engineering applications, i.e. from high-temperature electronics to superabrasive industrial tooling for cutting, drilling, polishing and shaping of ferrous alloys and hard ceramics, petroleum extraction, etc. (see Ref. 3 and references therein).

As it was found for cubic $BC_2N$ [1,4], diamond [5,6] and diamond-like $BC_5$ [7], creation of nanostructures by extreme pressure–temperature conditions leads to significant increase of the material hardness, mainly due to the Hall–Petch effect (nanosize effect which restricts dislocation propagation through the material) [8.9]. The first attempt to synthesize nanocrystalline cBN by direct conversion of graphite-like boron nitride at 18 GPa and 1900 K resulted in the formation of ultra-hard aggregated BN nanocomposite containing both cubic and wurzitic (wBN) polymorphs [10]. This nanostructured material shows very high Vickers hardness ($H_V$ up to 85 GPa) but relatively low thermal stability caused by the presence of metastable wBN. Successful synthesis of single-phase nanocrystalline cBN was performed by Solozhenko et al. [11] by direct solid-state phase transformation of graphite-like BN with "ideal random layer" (turbostratic) structure at 20 GPa and 1770 K. The material shows very high hardness ($H_V = 85(3)$ GPa), superior fracture toughness ($K_{Ic} = 10.5$ MPa·m$^{1/2}$), and high thermal stability and oxidation resistance (up to 1500 K).

Later synthesis of ultra-hard cBN-based nanostructured materials has been reported by other research groups [12-16], however, extremely high Vickers hardness claimed by some authors (up to 108 GPa for so called "nanotwinned" cBN [12,16]) is unconvincing [17,18]. It should be noted that data on hardness and elastic properties of nanocrystalline cBN are limited and rather controversial. In the present paper we report the mechanical properties of ultra-hard nanocrystalline cubic boron nitride from *in situ* study by nanoindentation technique, as well as from conventional microhardness and fracture toughness measurements.

---
[*] vladimir.solozhenko@univ-paris13.fr

**Experimental**

High-purity bulk nanocrystalline cubic boron nitride has been synthesized in a LPR 1000-400/50 Voggenreiter press with Walker-type module at 20 GPa and 1770 K by direct phase transformation of turbostratic graphite−like BN (tBN) following the method described previously [11]. According to X-ray diffraction study (TEXT 3000 Inel, CuKα1 radiation), the recovered bulks contain well-crystallized single-phase nanocrystalline cBN with lattice parameter $a = 3.616(1)$ Å, in perfect agreement with literature data [11]. Raman spectrum of the material shows strong broad band centered at ~400 cm$^{-1}$ and three week broad bands at ~820, ~1050 and ~1300 cm$^{-1}$ that is a fingerprint of nanocrystalline cBN with grain size less than 100 nm [11].

The recovered samples (cylinders 1.5–2 mm in diameter and 3-mm height) were hot mounted in electrically conductive carbon-fiber reinforced resin, and were polished with 9-μm and 1-μm diamond suspensions. Mechanical polishing was followed by vibropolishing with 0.04-μm SiO$_2$ colloidal solution for 24 hours. Such extensive polishing duration ensured the minimal sample surface damage that is required for accurate nanoindentation and microhardness measurements.

Microstructure characterization of nanocrystalline cBN has been performed on JEOL 3000F transmission electron microscope (TEM) which was used for scanning (STEM) and dark field (DFTEM) imaging and selected area electron diffraction (SAED). FEI NanoLab 600 dual beam scanning electron microscope (SEM) was used for the preparation of TEM lamella by focused ion beam (FIB) lift-out technique [19]. The same SEM was used for cross-validation of the indentation data from optical microscope. Atomic force microscope (AFM) Dimension 3100 Digital Instruments was used in tapping mode for characterization of Knoop and nanoindentation imprints, the later for pile-up correction.

Nanoindentation study has been performed on Micro Materials NanoTest Vantage system with trigonal Berkovich diamond indenter (the tip radius of 120 nm). The maximal applied load was 1000 mN. Loading at the rate of 0.5 mN/s was followed by a 10 s holding and unloading at the same rate.

Evaluation of the hardness and elastic modulus was performed in accordance to the Oliver-Pharr method [20]. The nanohardness was determined by Eq. 1:

$$H_N = \frac{P_{max}}{A(h_c)} \tag{1}$$

where $P_{max}$ is the maximum applied load and $A(h_c)$ is the projected contact area. The area function $A(h_c)$ was calibrated on a standard fused silica reference sample. Correction of the area function for the pile-up effects was based on the imprint topography data obtained on the actual samples by atomic force microscopy. The elastic recovery was estimated as the ratio of elastic work to the total work of the indentation by Eq. 2:

$$R_W = \frac{W_e}{W_p + W_e} \times 100\%$$ (2)

where $W_p$ and $W_e$ are plastic and elastic works, respectively. Reduced modulus $E_r$ was determined from stiffness measurements that are governed by elastic properties of the sample and diamond indenter via Eq. 3:

$$E_r = \left( \frac{1 - v_s^2}{E_s} + \frac{1 - v_i^2}{E_i} \right)^{-1}$$ (3)

where $E_s$, $E_i$ are Young's moduli and $v_s$, $v_i$ are the Poisson's ratios of the sample and indenter, respectively. The elastic modulus of material can be calculated for known properties of diamond ($E_i = 1141$ GPa and $v_i = 0.07$ [20]) and Poisson's ratio of the sample. Using the relation between Young's ($E$) and shear ($G$) moduli of an isotropic material

$$G = \frac{E}{2(1 + v)}$$ (4)

the shear modulus can be evaluated for the known value of Poisson's ratio.

Microhardness measurements have been performed using Ernst Leitz Wetzlar indentation tester under loads ranging from 0.25 to 5.0 N at 15 seconds dwell time. At least five indentations have been made at each load. The indentation imprints were measured with a Leica DMRME optical microscope under 1000× magnification in the phase contrast regime. The value of Knoop hardness ($H_K$) was determined by Eq. 5:

$$H_K = \frac{P}{0.070279 \cdot d^2}$$ (5)

where $P$ is the applied load and $d$ is the length of a large diagonal of an imprint.

Material hardness characterization by Vickers microindentation was omitted due to a poor accuracy related to the high elastic recovery of nanocrystalline cBN [13] and strong hardness overestimation as a result of low precision of imprint diagonal measurements (see Fig. 5b). However, Vickers indentation under 6 N and 7 N loads was used for characterization of material indentation fracture toughness ($K_{Ic}$), with at least 5 indentations at each load. The lengths of radial cracks emanating from the indent corners were measured by Leica DMRME optical and FEI NanoLab 600 SEM microscopes. The $K_{Ic}$ value was determined in terms of the indentation load $P$ and the mean length (surface tip-to-tip length $2c$) of the radial cracks according to Eq. 6 [21]:

$$K_{Ic} = x_v \cdot (E/H_V)^{0.5} \, (P/c^{1.5})$$ (6)

where $x_v = 0.016(4)$, $E$ is Young's modulus and $H_V$ is load-independent Vickers hardness. The latter value was taken as $H_V = 85$ GPa based on data reported earlier for the identical nanocrystalline cBN material [11].

Indentation fracture toughness was also assessed based on microindentation with Berkovich pyramid performed at 20 N load. Standardized form of Laugier $K_{Ic}$ formulation [22] was used which relates indentation load $P$ with indentation crack geometry parameters (imprint tip-to-corner length $a$, radial crack length $l$, and imprint-tip-to-crack-tip distance $c$) according to Eq. 7:

$$K_{Ic} = x_b \cdot (a/l)^{0.5} (E/H_N)^{1.5} P/c^{1.5} \tag{7}$$

where $x_b = 0.016(1)$ is the value of the constant corrected for three-sided Berkovich pyramid as opposed to the four-sided Vickers one [22], $E$ is Young's modulus, and $H_N$ is load-independent hardness for Berkovich indentation.

**Results and Discussion**

Fig. 1a depicts a typical nano-cBN microstructure as detected by scanning transmission microcopy in low-angle annular dark field (STEM LAADF) mode for grain diffraction contrast. The structure is characterized by non-equilibrium grain boundaries with only a few individual grains with equiaxed boundaries that underwent recrystallization. Presence of nano-twinning within all cBN grains, including those undergoing recrystallization, is also clearly visible.

Reported grain size distribution (Fig. 1b) is determined through the equivalent circle diameter method according to ASTM:E1382-97 [23] by considering the complete grain area determined by image processing algorithms for a series of STEM LAADF images. It can be seen that the grain size ranges from 10 to mainly 50 nm, with a few larger individual grains. The average value makes 35 nm.

Selected area diffraction pattern (Fig. 1c) shows a substantial deviation of individual reflections from the theoretical *111*, *220* and *311* diffraction rings which is indicative of numerous structural defects and residual strains as a result of the direct solid-state tBN-to-cBN transformation accompanied by a substantial volume change. Asterisk shows weak broad asymmetric wBN-like reflection caused by stacking faults, typical for displacive phase transformations [11].

Fig. 1d depicts an image composed of three superimposed dark field TEM images taken with large diffraction plane aperture of nearly 5 reciprocal nanometer diameter, as a way to include all *111* reflections. Data confirm nano-twinning of cubic boron nitride, absence of equiaxed grain boundaries, while diffuse appearance of the grains also confirms presence of structural defects.

From 18 independent nanoindentation experiments it was found that in the whole studied range of peak indentation loads (80-1000 mN) the measured nanohardness of bulk nanocrystalline cBN is almost constant and makes $H_N = 78(2)$ GPa. Fig. 2 shows the characteristic load-displacement curves. The elastic recovery of nano-cBN has been estimated by Eq. 2 as 79(2)% that is much higher than elastic recovery of single-crystal cubic BN (60% [24]). The Young's modulus of $E = 961(39)$ GPa was calculated by Eq. 3 using the experimental value $E_r = 524(9)$ GPa and Poisson's ratio $\nu = 0.07$ estimated from the experimental value of nano-cBN bulk modulus (375(4) GPa [25]) according to the relation between bulk and Young's moduli:

$$B = \frac{E}{3(1-2\nu)} \qquad (8)$$

Thus, Young's modulus of nano-cBN is noticeably higher than 909 GPa value calculated from the elastic stiffness constants of single-crystal cBN [26]. The shear modulus of nano-cBN was evaluated by Eq. 4 as $G = 449(18)$ GPa which is 12% higher that the maximal value $G = 402$ GPa reported for translucent polycrystalline cBN with grain size of 2-4 μm [27].

The measured Knoop hardness of bulk nanocrystalline cubic boron nitride decreases with the load and at 5 N reaches the asymptotic value of $H_K = 63(2)$ GPa (Fig. 3) which is higher than previously reported Knoop hardness values for cBN-based nanostructured bulk materials [13,15,16,28,29].

Atomic force microscopy data on Knoop imprint (Fig. 4) indicate that, similarly to earlier reported observations [13], the actual length of the short diagonal $d_s$ is smaller than the value prescribed by the indenter geometry [30] due to material elastic recovery. Bearing in mind that Knoop hardness is the ratio of testing load to the projected area of the indentation [30], such reduction in $d_s$ value (and respective reduction of the projected area) results in an increase of hardness. AFM data indicate that 7 % reduction of projected area took place which should lead to the proportional increase in $H_K$ value as determined based on the long diagonal $d$ (Eq. 5).

It can be seen that Knoop indentation does not lead to formation of radial cracks as in the case of Vickers (Fig. 5a) and Berkovich (Fig. 6) pyramids, thus making it a reliable technique for characterizing only the material hardness, as opposed to combined measure of hardness and fracture resistance for the two other indentation techniques.

Indentation fracture toughness at both indentation loads of 6 N and 7 N (Fig. 5a) demonstrated that the minimum requirement criterion $c \geq 2a$ for crack length $c$ and half-imprint $a$ is fulfilled ($c/a = 3.6$ at 6 N and $c/a = 3.8$ at 7 N). Close-up observation of post-indentation imprint (Fig. 5b) indicates that an additional source of uncertainly is related to locating the corners of Vickers imprint which is disguised by the material elastic recovery and a series of concentric cracks that are formed parallel to the indenter facets. This finding verifies low accuracy of Vickers hardness measurement [13] which may lead to significant hardness overestimations, as it is reciprocal to diagonal squared. Apart from that, it is visible that indentation cracks initiate not directly at the corner but some distance inside the residual imprint. Both factors might contribute to the underestimation of fracture toughness.

The average fracture toughness of nano-cBN is practically the same for both loads and is estimated as $K_{Ic} = 5.0 \pm 0.3$ MPa·m$^{1/2}$ at 7 N which is almost twice higher than the 2.8 MPa·m$^{1/2}$ value for single-crystal cBN [31], yet this value is significantly lower than the one reported earlier for another nano-cBN materials of the same class [11]. Presence of structural defects and residual strain (see Fig. 1) which contributes to anomalous hardness increase are most likely compromising the fracture toughness.

The average fracture toughness determined for indentation crack geometry by Berkovich pyramid (Fig. 6) was estimated as $K_{Ic} = 4.7\pm0.4$ MPa·m$^{1/2}$. This value is only slightly lower than the one for indentation fracture toughness with conventional Vickers pyramid.

**Conclusions**

The data on mechanical and elastic properties of nanocrystalline cubic boron nitride are summarized in the Table together with the corresponding values for single-crystal and microcrystalline cBN. Due to extremely high hardness and elastic recovery as well as high thermal and chemical stability [11], the synthesized bulk nanocrystalline cubic boron nitride offers further substantial improvement and promise as an exceptional superabrasive for a wide range of engineering applications, including established areas for existing conventional nano-cBN solutions, such as micro- and nano-machining of ferrous alloys for die and mold applications [33], and use for manufacture nanoindentation tips for operation at elevated temperature [34].

**Acknowledgements**

The authors thank Dr. Kirill A. Cherednichenko for assistance in high-pressure synthesis of nano-cBN and Dr. Filip Lenrick for TEM study. This work was financially supported by the European Union's Horizon 2020 Research and Innovation Programme under the Flintstone2020 project (grant agreement No 689279).

Table   Hardness, elastic moduli, Poisson's ratio and fracture toughness of cubic boron nitride
        (data of the present work are given in bold)

| | $H_V$ | $H_K$ | $H_N$ | $E$ | $G$ | $B$ | $\nu$ | $K_{Ic}$ |
|---|---|---|---|---|---|---|---|---|
| | GPa | | | | | | | MPa·m$^{1/2}$ |
| *nano*-cBN | 85 [11] | **63(2)** | **78(2)** | **961(39)** | **449(18)** | 375 [25] | **0.07** | **4.9(4)** |
| cBN | 62 [24]* | 44 [24]* | 55 [24]* | 909 [24] | 407$^\dagger$ | 392 [32]* | 0.12 [27] | 2.8 [31]* |

\* Single crystal

$^\dagger$ Calculated using Eq. 4

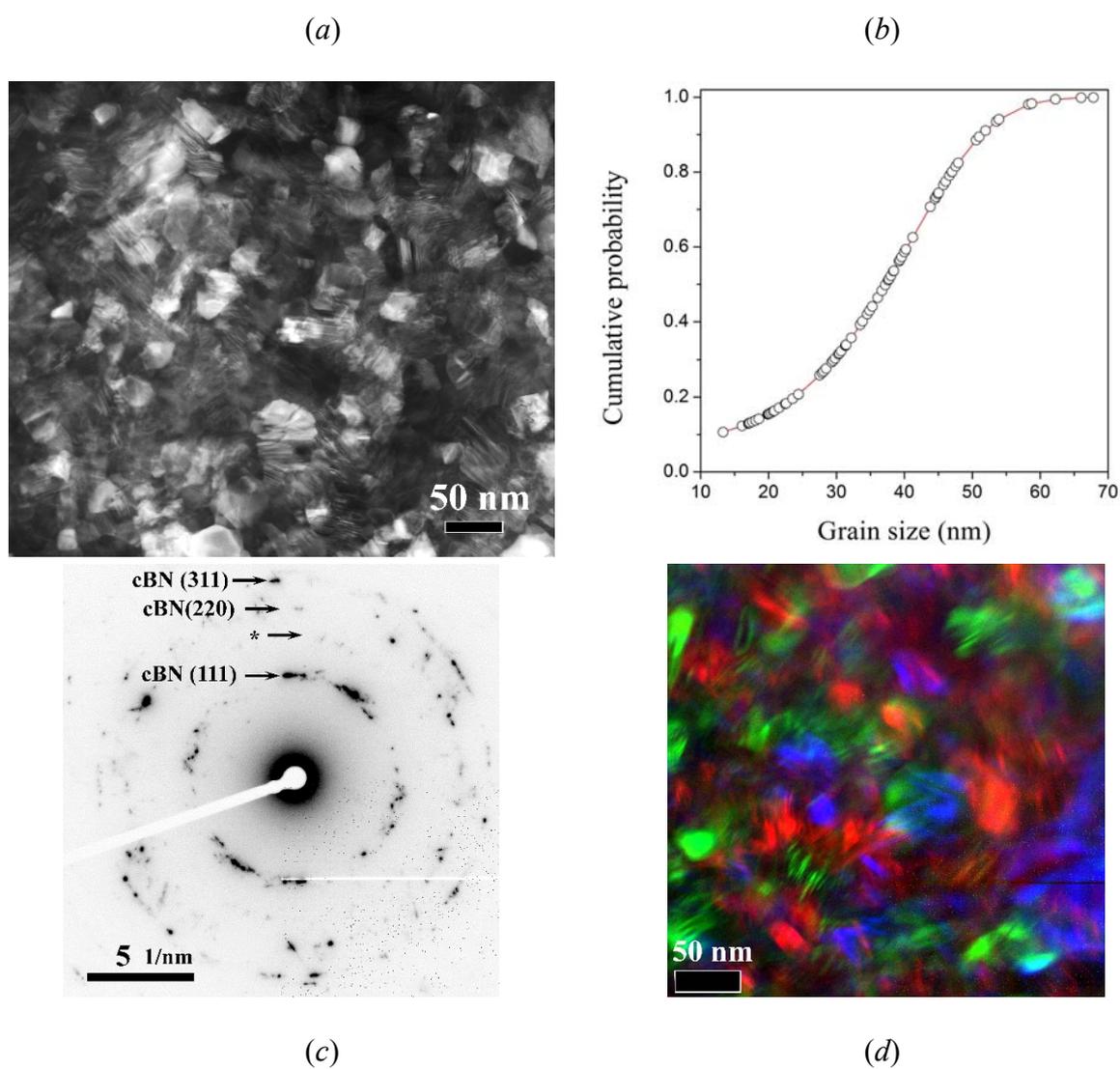

(*c*)       (*d*)

Fig. 1 (*a*) STEM LAADF image and (*b*) cumulative density function of grain-size distribution of bulk nanocrystalline cubic boron nitride; (*c*) (inverted contrast) SAED pattern from the TEM lamella, and (*d*) three superimposed false color dark field TEM images.

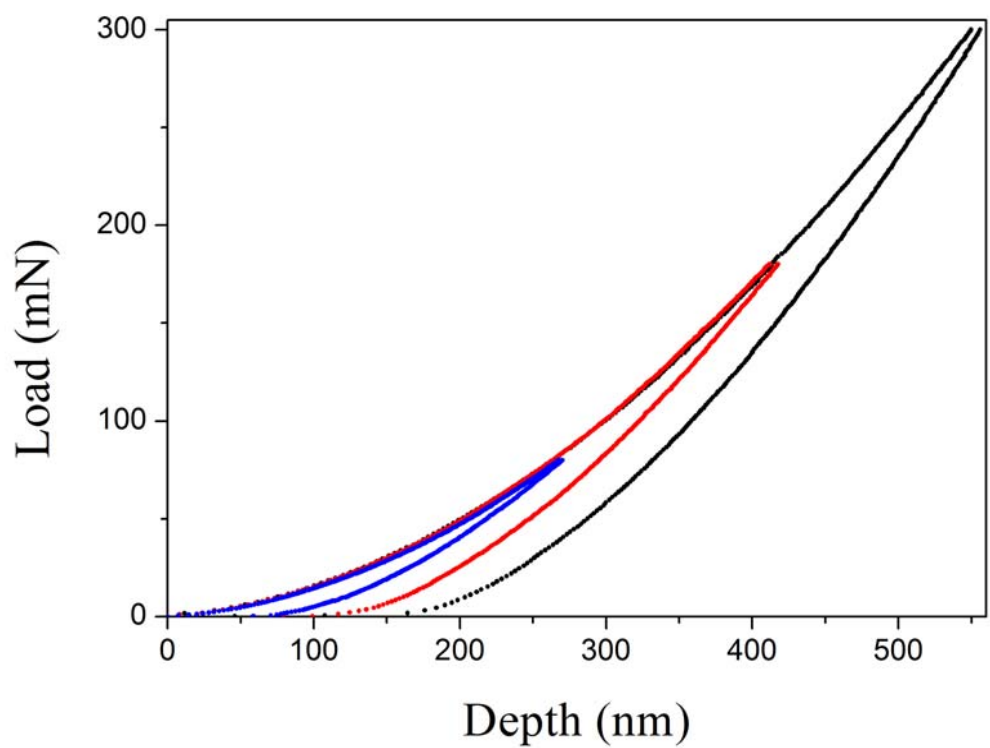

Fig. 2    Load-displacement curves of bulk nanocrystalline cubic boron nitride at 80 mN, 180 mN and 300 mN peak indentation loads.

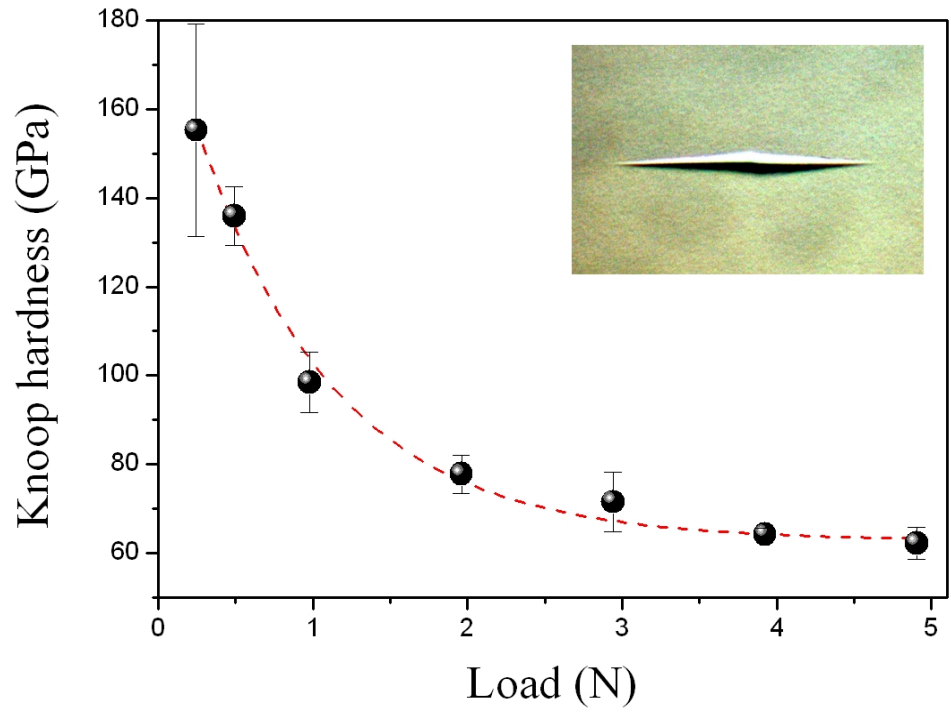

Fig. 3    Knoop hardness of bulk nanocrystalline cubic boron nitride *vs* load.
Inset: optical microscope image of the imprint at 5 N load.

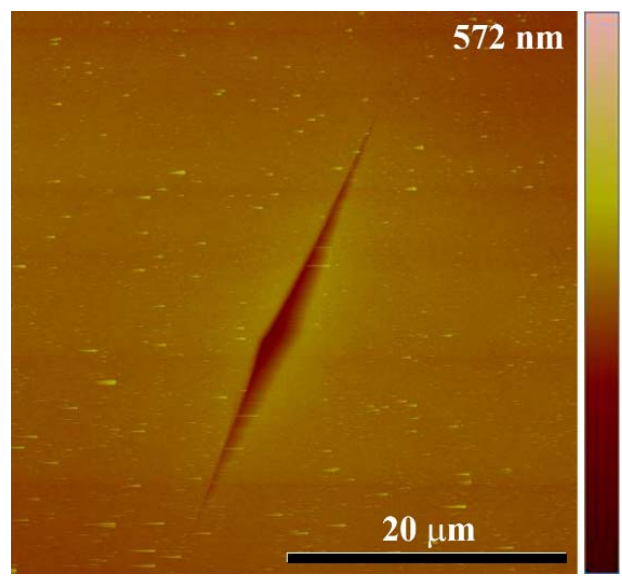

Fig. 4.  AFM image of Knoop imprint at 5 N load.

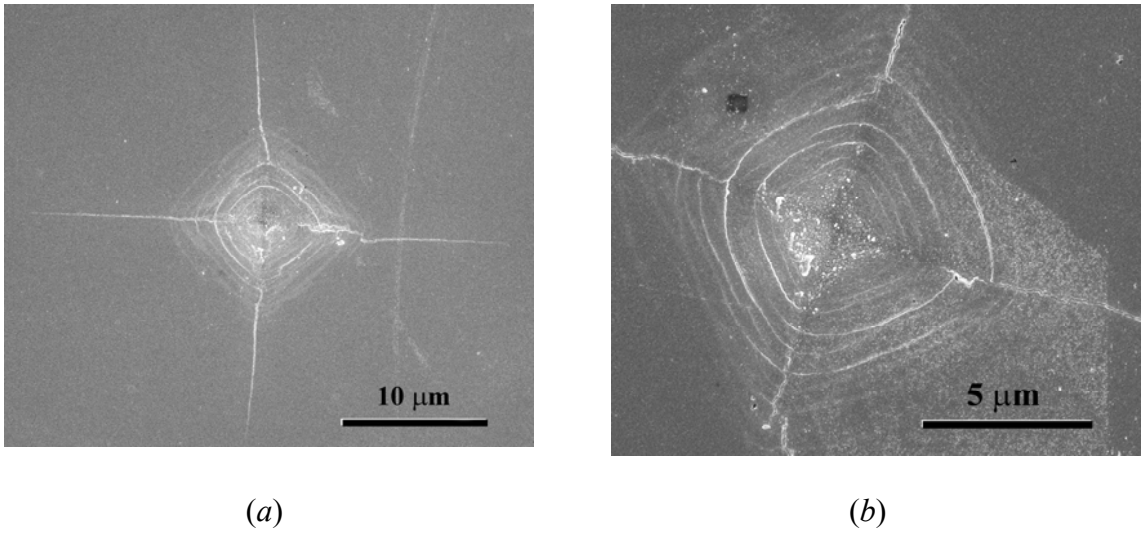

(*a*)                                    (*b*)

Fig. 5    (*a*) SEM image of Vickers imprint after indentation fracture toughness test (6 N load), and (*b*) close-up SEM image of Vickers imprint at 7 N load.

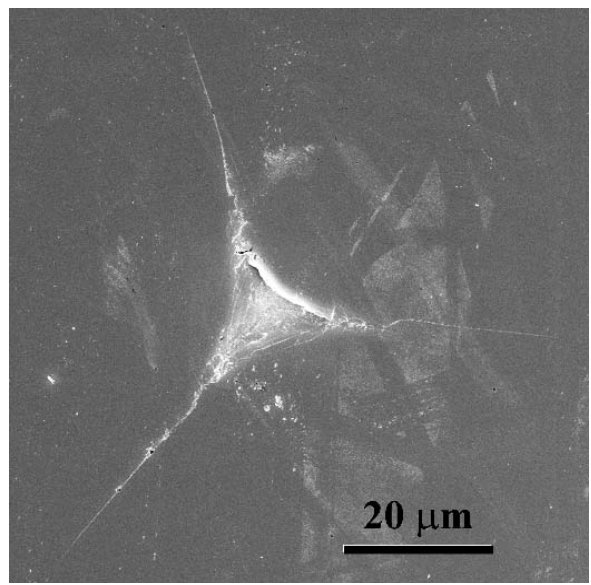

Fig. 6   SEM image of Berkovich imprint after microhardness test at 20 N load.